\documentclass[12pt]{article}
\usepackage{epsfig} \usepackage{amssymb} \usepackage{amsfonts}
\def\eV{{\rm e\kern-0.12em V}} \def\GeV{{\rm G}\eV} \def\MeV{{\rm M}\eV}
\def\msbar{\relax\ifmmode\overline{\rm MS}\else{$\overline{\rm MS}${ }}\fi}
\def\alphan{\relax\ifmmode{\alpha_{\rm an}}\else{$\,\alpha_{\rm an}${ }}\fi}
\def\alphant{\relax\ifmmode{\alpha_{\rm an}(M_{\tau})}\else{$\alphan(M_{\tau})${ }}\fi}
\def\albars{\relax\ifmmode{\bar{\alpha}_s}\else{$\,\bar{\alpha}_s${ }}\fi}
\def\albarsZ{\relax\ifmmode{\bar{\alpha}_s}(M_Z)\else{$\,\albars(M_Z)${ }}\fi}
\def\albarsQ{\relax\ifmmode{\bar{\alpha}_s(Q^2)}\else{$\albars(Q^2)${ }}\fi}
\def\albarsq{\relax\ifmmode{\bar{\alpha}_s(q^2)}\else{$\albars(q^2)${ }}\fi}
\def\asmz{\relax\ifmmode\bar \alpha_s(M_Z^2)\else{$\albars(M_Z^2)${ }}\fi}
\def\asmzs{\relax\ifmmode{\albars(M_z^2)}\else{{$\albars(m_z^2)$}{ }}\fi}
\def\tildal{\relax\ifmmode{\tilde{\alpha}}\else{$\,\tilde{\alpha}${ }}\fi}
\def\tildalt{\relax\ifmmode{\tildal(M_{\tau})}\else{$\tildal(M_{\tau})${ }}\fi}
\def\tildals{\relax\ifmmode{\tilde{\alpha}(s)}\else{$\tilde{\alpha}(s)${ }}\fi}
\def\asq{\relax\ifmmode\albarsq\else{$\,\albarsq${ }}\fi}
\def\as{\relax\ifmmode\alpha_s\else{$\,\alpha_s${ }}\fi}
\def\agoth{\relax\ifmmode{\mathfrak A}\else{$\,{\mathfrak A}${ }}\fi}
\def\pisq{\relax\ifmmode{\pi^2}\else{${\pi^2}${ }}\fi}
\def\agothk{\relax\ifmmode{\mathfrak A}_k\else{${\mathfrak A}_k${ }}\fi}
\def\acal{\relax\ifmmode{\cal A}\else{${\,\cal A}${ }}\fi}
\def\acalk{\relax\ifmmode{\cal A}_k\else{$\,{\cal A}_k${ }}\fi}
\newcommand{\beglab}{\begin{equation}\label}
\newcommand{\beq}{\begin{equation}} \newcommand{\eeq}{\end{equation}}
\begin{document}
\begin{flushright} JINR Preprint E2--2001--153; to be published in EJP C
\end{flushright} \medskip

\begin{center}
{\large\sf Analytic Perturbation Theory in analyzing
some QCD observables } \\
\bigskip

D.V. Shirkov \\
{\it Bogoliubov Laboratory, JINR, 141980 Dubna, Russia\\
e-address: shirkovd@thsun1.jinr.ru} \end{center} \smallskip

\centerline{\bf Abstract} {\small The paper is devoted to application of
recently devised ghost--free Analytic Perturbation Theory (APT) for
analysis of some QCD observables. We start with the discussion of the main
problem of the perturbative QCD --- ghost singularities and with the resume
of its resolving within the APT. \par

  By a few examples in the various energy and momentum transfer regions
(with the flavor number $f=3,4$ and 5) we demonstrate the effect of
improved convergence of the APT modified perturbative QCD expansion. \par
  Our first observation is that in the APT analysis the three-loop
contribution ($\sim\,\as^3$) is as a rule numerically inessential.
This gives raise a hope for practical solution of the well--known problem
of asymptotic nature of common QFT perturbation series. \par
  The second result is that a usual perturbative analysis of time-like
events with the big $\pi^2$ term in the $\alpha_s^3$ coefficient is not
adequate at $s\leq 2\:\GeV^2\,.$ In particular, this relates to
$\tau$ decay. \par

 Then, for the ``high" ($f=5$) region it is shown that the common two-loop
(NLO, NLLA) perturbation approximation widely used there (at
$10\:\,\GeV\lesssim\sqrt{s}\lesssim 170\:\,\GeV$) for analysis
of shape/events data contains a systematic negative error of a 1--2 per
cent level for the extracted $\albars^{(2)}\,$ values. \par
Our physical conclusion is that the \asmz value averaged over the
$f=5\,$ data appreciably differs $\,<\asmz>_{f=5}\, \simeq 0.124\,$
from the currently accepted ``world average" $(=0.118\,)$.}

\section{\sf Preamble}
In QCD, a dominant means of theoretical analysis is based on perturbation
power expansion supported by appropriate renormalization group (RG)
summation. This perturbative QCD (pQCD) satisfactory correlates the bulk
of experimental data in spite of the fact that the RG invariant power
expansion parameter \albars is not a ``small enough" quantity. Nowadays,
the physically accessible region corresponds to three, four and five
$(f=3,4,5)\,$ flavor numbers (of active quarks). Just in the three--flavor
region there lie unphysical singularities of the central theoretical object
--- invariant effective coupling \albars. These singularities, associated
with the QCD scale parameter $\,\Lambda_{f=3}\simeq 400\,$ \MeV, complicate
theoretical interpretation of data in the ``small energy" and ``small
momentum transfer" regions $(\,\sqrt{s},\,q\equiv\sqrt{q^2}\lesssim 1
\div 1.5 \: \GeV).$ On the other hand, as it is well known, their
existence contradicts some general statements of the local QFT. \par

 In this paper, we first discuss this main problem of the pQCD, the
singularities lying in the physically accessible domain, and then give
a resume of its solution within the recently devised ghost--free Analytic
Perturbation Theory (APT) that resolves the problem without using any
additional adjustable parameters. Then, we give some impressive results of
the APT application for analysis of QCD observables.

\subsection{Invariant QCD coupling and observables}
  Usually, the perturbative QCD part of theoretical contribution
to observables in both the space-- and time--like channels is presented
in the form of two-- or three--term power expansion
 \beq \label{rnew}
\frac{O(x)}{O_0}=1+r(x)\,;\quad r(x)=c_1\,\albars(x)+c_2\,\albars^2+
c_3\,\albars^3 +\dots \,; \quad x=q^2 \,\;\, \mbox{or}\, =s \eeq
(our coefficients are normalized $c_k=C_k\,\pi^{-k}$ differently from the
commonly adopted $C_k\,$, like in Refs.\cite{pdg00,beth00, bardin}) over
powers of the effective QCD coupling \albars which is supposed {\it ad hoc\,}
to be of the same form as in both the channels, e.g., in the massless
three--loop case
\begin{eqnarray}\label{als3}
\albars^{(3)}(x)&=&\frac{1}{\beta_0L}-\frac{b_1}{\beta_0^2}\frac{\ln L}
{L^2} +\frac{1}{\beta_0^3L^3}\left[b_1^2(\ln^2L-\ln L-1)+b_2\right]+ \,
\nonumber \\
 &+&\frac{1}{\beta_0^4L^4}\left[b_1^3\left(-\ln^3L+\frac{5}{2}\ln^2L+
2\ln L-\frac{1}{2}\right)- 3b_1b_2\ln L+\frac{b_3}{2}\right]\,.
\end{eqnarray}

Here, $L=\ln (x/\Lambda^2)\,,$ and for the beta-function coefficients we use
the normalization
$$
\beta(\alpha)=-\beta_0\,\alpha^2-\beta_1\,\alpha^3-\beta_2\,\alpha^4+\dots
=-\beta_0\,\alpha^2\left(1+b_1\,\alpha +b_2\,\alpha^2+\dots\right)\,,$$
that is also free of the $\,\pi\,$ powers. Numerically, they are of an
order of unity {\small
$$\beta_0(f)=\frac{33-2\,f}{12\pi}\,;~\ b_1(f)=\frac{153-19f}{2\pi(33-2f)}
\,;\,\, \beta_0(4\pm 1)=0.875 \pm 0.005\,;\,\,
  b_1(4\pm 1)=0.490^{-0.089}_{+0.076}\,. $$}  \smallskip

 Meanwhile, the RG notion of invariant coupling was first introduced in QED
\cite{rg56} in the space-like region in terms of a real constant $\,z_3\,$
of the finite Dyson renormalization transformation. Just this QED Euclidean
invariant charge $\,\bar{e}(q)\,$ is the Fourier transform of the space
distribution $\,e(r)\,$ of electric charge (arising due to vacuum
fluctuations around a point ``bare" electron) discussed by Dirac
\cite{dirac34} in 30s --- see Appendix IX in the textbook \cite{qf}. \par

  Generally, in the RG formalism (for details, see, e.g., the chapter
``Renormalization group" in the monograph \cite{kniga} and/or Section 1 in
Ref.\cite{DSh95}) the notion of invariant coupling $\,\bar{g}(q)\,$ is
~{\it defined only in the space--like domain}.

 In particular, this means that if some observable $O(q^2)$ is physically
a function of one kinematic Lorentz--invariant space--like argument
$q^2\,,$ then, due to its renormalization invariance, it should be a
function of RG invariants only. E.g., in the one--coupling massless case
$$
O(q^2/\mu^2,\,g_{\mu})=F\left(\bar{g}(q^2/\mu^2,\,g_{\mu})\right)\quad
\, \mbox{with} \,\,\quad F(g)=O(1, g)\,.$$
 Due to this important property, in the weak coupling case we deal with
the {\it functional expansion\,} of an observable $O(q^2)\,$ in powers of
invariant coupling $\bar{g}\,.$
This is a real foundation of QCD power expansion (\ref{rnew}) in the
Euclidean case with $\,x=q^2\,.$ At the same time, inside the RG formalism,
there is no natural means for defining invariant coupling
$\,\,\tilde{g}(s)\,;\,s=-q^2$ and perturbative expansion for an observable
$\,\tilde{O}(s)\,$ in the time--like region. \par \smallskip

  Nevertheless, in modern practice, people commonly use the same singular
expression for the QCD effective coupling \albars, like (\ref{als3}), in
both the space-- and time--like domains.
 The only price usually paid for this transferring from the Euclidean to
Minkowskian region is the change of numerical expansion coefficients.
The time--like ones $\,c_{k\geq 3}=d_k-\delta_k\,$ include negative
``$\pi^2$ terms"  proportional to $\pi^2$ and lower expansion coefficients
$c_k$ 
\begin{equation}\label{deltas}
\delta_3=\frac{(\pi\beta_0(f))^2}{3}\,c_1\,,~~\delta_4=(\pi\beta_0)^2
\,(c_2+\frac{5}{6}\:b_1\,c_1)\, \dots\,\,.\eeq 
These (rather essential, as far as $\pi^2\beta_0^2(f=4\pm 1)=
4.340^{-.666}_{+.723}$) structures $\,\delta_k\,$ arise
\cite{rad82} --- \cite{kat95} in the course of analytic continuation
from the Euclidean to Minkowskian region. The coefficients $d_k\,$ should be
treated as genuine $k\,$th--order ones. Just they are calculated via
 the relevant Feynman diagrams. \smallskip

\begin{minipage}[]{155mm}
\begin{center} {\sf\large Table 1}  \label{tab1}\smallskip

\centerline{\sf Minkowskian $c_k$ and  Euclidean $d_i=c_i+\delta_i$
expansion coefficients and their differences.} \medskip

\begin{tabular}{|c|c||c|c|c|c|c|c|}  \hline
Process & f &$c_1=d_1$&$c_2=d_2$& $c_3$ &$d_3$&$\delta_3$&
$\delta_4$ \\ \hline\hline
$\tau$ decay& 3&$1/\pi$&.526&$0.852$& 1.389& 0.537  & 5.01 \\ \hline
$e^+e^-$    & 4 &.318 &.155 &-\,0.351& 0.111 & 0.462 & 2.451 \\ \hline
$e^+e^-$    & 5 &.318 &.143 &-\,0.413 &-\,0.023& 0.390  & 1.752 \\ \hline
$Z_0$ decay & 5 &.318&.095 &-\,0.483 &-\,0.094& 0.390  & 1.576  \\ \hline
\end{tabular}\end{center}
\end{minipage} \medskip

 To demonstrate the importance of the~``$\pi^2$ terms", we took the
$\,f=3\,$ case for $\tau$--decay, the $f=4\,,5\,$ cases for $e^+e^-\to$
hadron annihilation and the $Z_0$ decay (with $f=5$) --- see Table 1 in
which we also give values for the $\pi^2$--terms\,. In the normalization
(\ref{rnew}), all coefficients $\:c_k\,,\: d_k\,$ and $\:\delta_{k}\,$
are of an order of unity. In the $\,f=4,5\,$ region the contribution
$\:\delta_3\,$ prevails in $\:c_3\,$ and $|d_3| \ll |c_3|\,$ (see also
Table II in Bjorken's review \cite{bjork89}). \medskip

\subsection{Unphysical singularities}
Let us remind to the reader that the ghost-trouble first discovered in
QED in the mid-50s (and quite soon in the renormalizable version of pion-nucleon
interaction) was considered there as a serious argument in favor of inner
inconsistency of the whole local QFT. In the QED case, the ghost singularity
lies far above the mass of the Universe and has no pragmatic meaning. \par
  However, in QCD it lies in the quite physical infrared (IR) region and we
are forced to face it without any excusing arguments. This means that, if
one believes in QCD as in a consistent, physically important theory, one
has no other possibility as to consider the QCD unphysical singularities
as an artefact of some approximations used in pQCD. This point of view is
supported by some lattice simulations and solution of Schwinger--Dyson
equations --- see, e.g., Section 5.3. in a recent review \cite{alkofer01}.

For illustration of the fundamental inconsistency of current pQCD
practice connected with unphysical singularities, take the well-known
relation between the so--called Adler function $D$ and the total
cross--section ratio $R$ of a related process
\begin{equation}\label{Adler}
 D(q^2) = q^2\int^{\infty}_0 \frac{R(s)\,d s}{(s+q^2)^2}\,.\eeq

 In the case of inclusive $e^+e^-$ annihilation into hadrons, $R(s)\,$ is
the ratio of cross--sections presented in the form $R(s)= 1+ r(s)\,$ with a
function $r$ expandable in powers of $\albars(s)\,$ like in Eq.(\ref{rnew}).
At the same time, the Adler function is also used to be presented in the
form $\,D=1+d\,$ with $d$ expanded in powers of \asq. \par
  Here, we face two paradoxes. First, \asq  and, hence, the perturbative
$\,D(q^2)\,$ obeys --- see eq.(\ref{als3}) --- non-physical singularity at
$\;q^2 =\Lambda^2\;$ in evident contradiction with representation
(\ref{Adler}). Second, the integrand $R(s)\,,$ being expressed via powers
of $\albars(s)\,,$ obeys non-integrable singularities at $\;s=\Lambda^2\;,$
which makes the r.h.s. of eq.(\ref{Adler}) senseless. \par
  This second problem is typical of inclusive cross--sections, e.g., for
the $\tau$ hadronic decay. Generally, in the current literature it is
treated in a very strange way --- by shifting the contour of integration
from the real axis with strong singularities on it into a complex plane.
However, such a ``physical" trick cannot be justified within the theory
of complex variable.

\subsection{The ``ghost" problem resolving}
  Meanwhile, as it is known from the early 80s, the perturbation
representation (\ref{rnew}) for the Minkowskian observable with the
coefficients modified by the $\pi^2$--terms is valid only at a small
parameter $\pi^2/\ln^2(s/\Lambda^2)\,$ values, that is in the region of
sufficiently high energies
$\,W\equiv\sqrt{s}\gg\Lambda e^{\pi/2}\simeq 2\,\GeV\,$. \par
 Here, it is appropriate to remind the construction devised by Radyushkin
\cite{rad82} and Kras\-ni\-kov---Pivovarov \cite{kras82} (RKP procedure)
about 20 years ago. These authors used the integral transformation
\beglab{R-oper}
R(s)=\frac{i}{2\pi}\,\int^{s+i\varepsilon}_{s-i\varepsilon}\frac{d
z}{z}\, D_{\rm pt}(-z)\equiv{\bf R}\left[D_{\rm }(q^2)\right]\,\eeq
reverse to the Adler relation (\ref{Adler}) (that is treated now as
integral transformation)
\begin{equation}\label{D-trans}
R(s) \to D(q^2) = q^2\int^{\infty}_0 \frac{\,R(s)\,d s}{(s+q^2)^2}\,
\equiv {\cal\bf D} \left\{ R(s)\right\}\, \eeq
for defining modified expansion functions
\beglab{agot-def}
\agothk(s) = {\bf R}[\as^k(q^2)] \eeq
for the perturbative QCD contribution
\beglab{r-apt}
 r(s)= d_1 \agoth_1(s) + d_2\agoth_2(s)+d_3\agoth_3(s)\,\eeq
to an observable in the time--like region. \par
   At the one-loop level, with the effective coupling $\albars^{(1)}=
\left[\beta_0 \ln(q^2/\Lambda^2)\right]^{-1}$ one has
\beglab{tildal1}
\agoth_1^{(1)}(s)={\bf R}\left[\albars^{(1)}\right]\,= \frac{1}{\pi\beta_0}
\arccos\frac{L}{\sqrt{L^2+\pi^2}}=\frac{1}{\beta_0}\left[\frac{1}{2}-
\frac{1}{\pi}\arctan\frac{L}{\pi}\right]\,;\quad L=\ln\frac{s}{\Lambda^2}
\:\eeq
and for higher functions
\beq\label{agoth234-1}
\agoth_2^{(1)}(s)=\frac{1}{\beta_0^2\left[L^2+\pi^2\right]}\:; \quad
\agoth_3^{(1)}(s)=\frac{L}{\beta_0^3\left[L^2+\pi^2\right]^2}\:; \quad
\agoth_4^{(1)}(s)=\frac{L^2-\pi^2/3}{\beta_0^4\left[L^2+\pi^2\right]^3}
\;,\eeq
which {\it are not powers\,} of $\agoth_1^{(1)}(s)\,.$ \par
The r.h.s of (\ref{tildal1}) at $L\geq 0\,$ can also be presented in the
form
$$
\hspace{60mm}\agoth_1^{(1)}(s)=\frac{1}{\pi\beta_0}
\arctan\frac{\pi}{L}\, \hspace{55mm} (9a) $$
convenient for the UV analysis. Just this form (9a) was discovered in
the early 80s in Refs.\cite{schr2} and \cite{rad82}, while
eqs.(\ref{agoth234-1}) in Refs.\cite{rad82} and \cite{kras82}. All
these papers dealt with HE behavior and did not pay proper attention to
the region $L\leq 0\,.$ \par
 On the other hand, expression (\ref{tildal1}) was first discussed only
15 years later by Milton and Solovtsov \cite{ms97}. Just these authors
first made an important observation that expression (\ref{tildal1})
represents a continuous monotone function without unphysical singularity at
$L=0\,$ and proposed to use it as an effective ``Minkowskian QCD coupling"
$\,\tildal(s)\equiv\agoth_1(s)\,$ in the time--like region. \par
    For the two--loop case, to the popular approximation
$$\beta_0\bar{\alpha}_{s,pop}^{(2)}(q^2)=\frac{1}{l}-b_1(f)
\frac{\ln l}{l^2}\:; \quad l=\ln\frac{q^2}{\Lambda^2}\,$$
there corresponds \cite{rad82,brs00}
\beq\label{tildal2pop}
\tildal^{(2)}_{pop}(s)\equiv \agoth^{(2,pop)}_{1}(s)=\left(1+\frac{b_1 L}
{L^2+\pi^2}\right)\tildal^{(1)}(s) -\frac{b_1}{\beta_0}\frac{\ln\left[
\sqrt{L^2+\pi^2}\right]+1}{L^2+\pi^2}\,. \eeq

 At $L\gg \pi$, by expanding this expression and $\agoth_2$ from
(\ref{agoth234-1}) in powers of $\,\pi^2/L^2$ we arrive at the
$\pi^2$--terms (\ref{deltas}).

 Both the functions (\ref{tildal1}) and (\ref{tildal2pop}) are
monotonically decreasing with a finite IR value $\,\tildal(0)=
1/\beta_0(f=3)\simeq 1.4\,.$ They have no singularity at $L=0$. Higher
functions go to zero $\agothk(0)=0\,$ in the IR limit. \par

As it has first been noticed in \cite{tow00,tmp01}, by applying the
transformation ${\bf D}$ (\ref{D-trans}) to functions $\agothk(s)$, instead
of \asq powers, we obtain expressions ${\bf D}[\agothk(s)]=\acalk(q^2)\,$
that are also free of unphysical singularities. These functions have been
discussed at 90s \cite{rapid96} --- \cite{ss99tmp} in the context of the
so--called ``Analytic approach" to perturbative QCD. \par

   Therefore, this Analytic approach in the Euclidean region and the RKP
formulation for Minkowskian observables can be united in the single scheme,
the ``Analytic Perturbation Theory" --- APT, that has been formulated quite
recently in our papers \cite{tow00} and \cite{tmp01}. In the next Section,
we give a short resume of this APT construction and then, in Sections 3 and
4, present the results of its practical applications.\par

\section{\sf The APT --- a closed theoretical scheme\label{s2}}     
 The  APT scheme closely relates two ghost--free formulations of
modified perturbation expansion for observables.\par

\subsection{Relation between Euclidean and Minkowskian} The first one, that
was initiated in the early eighties \cite{rad82,kras82} and outlined above,
changes the standard power expansion (\ref{rnew}) in the time-like region
into the nonpower one (\ref{r-apt}). It uses operation Eq.(\ref{R-oper}),
that is reverse ${\bf R}=[{\bf D}]^{-1}\,$ to the one defined by the ``Adler
relation" (\ref{D-trans}) and transforms a real function $\,R(s)\,$ of a
positive (time--like) argument into a real function $\,D(q^2)\,$ of a
positive (space--like) argument.\par
  By operation ${\cal\bf R}\,,$ one can define \cite{ms97} the RG--invariant
Minkowskian coupling $\tildal(s)={\bf R}\left[\albars\right] \,,$ and its
``effective powers" (\ref{agot-def}) that are free of ghost singularities.
 Some examples are given by expressions (\ref{tildal1}), (\ref{agoth234-1})
and (\ref{tildal2pop}). At the one--loop level, they are related by the
differential recursion relation $\,k\beta_0\agoth_{k+1}^{(1)}=-(d/dL)
\agothk^{(1)}$ and {\it are not powers} of  $\,\agoth^{(1)}_1\,.$ \par
  By applying  ${\bf D}$ to $\agoth_k(s)\,,$ one can ``try to return"
to the Euclidean domain. However, instead of \as powers, we arrive at
some other functions
\beglab{acal}
\acalk(q^2)={\bf D}\left[\agothk \right]\,,\eeq
 analytic in the cut $q^2$-plane and free of ghost singularities.
  At the one--loop case
\beq\label{4} \beta_0 \acal^{(1)}_1(q^2) = \frac{1}{\ln (q^2/\Lambda^2)} -
\frac{\Lambda^2}{q^2-\Lambda^2}\,,\quad\beta_0^2 \acal^{(1)}_2(q^2)=
\frac{1}{\ln^2 (q^2/\Lambda^2)}+ \frac{q^2\Lambda^2}{(q^2-\Lambda^2)^2}\,,
\,\,\, \dots \,.\eeq

 These expressions have been originally obtained by other means
\cite{rapid96,prl97} in the mid--90s. The first function $\acal_1=
\alphan(q^2)\,,$ an analytic invariant Euclidean coupling, should now be
treated as a {\it counterpart\/} of the invariant Minkowskian coupling
$\tildal(s)=\agoth_1(s)\,.$ Both $\alphan$ and $\tildal$ are real
monotonically decreasing functions with the same maximum value
$$
\alphan(0)=\tildal(0)=1/\beta_0({f=3})\simeq1.4\,$$
in the IR limit\footnote{
Note that the transition from the usual invariant \msbar coupling \as to
the Minkowskian \tildal and Euclidean \alphan ones can be understood as a
transformation to new renormalization schemes. At the one--loop case
\beglab{altrans}
\as\to \tildal^{(1)}=\frac{1}{\pi\beta_0}\arctan(\pi\beta_0\as)\hspace{7mm}
\mbox{and}\,\hspace{6mm}\,\,\as\to\alphan^{(1)}=\as+
\frac{1}{\beta_0}\left(1-e^{1/\beta_0\as}\right)^{-1} \,. \eeq
 Here, the first transition looks  ``quite usual" as $\tildal$ can be
expanded in powers of \as, while the second one in the weak coupling case
behaves like the identity transformation as far as the second
nonperturbative term $~e^{-1/\beta_0\as}$ leaves no ``footsteps" in the
power expansion.\par
 For both $\tildal^{(1)}$ and $\alphan^{(1)}$ the corresponding  $\beta$
functions have zero at $\alpha=1/\beta_0$ and are symmetric under reflection
$\left[\alpha -1/2\beta_0\right]\to -\left[\alpha -1/2\beta_0\right]\,.$
Moreover, the $\beta$ function for $\tildal(s)\,$ turns out to be equal to
the spectral function for $\alphan(q^2)\,$ -- see below Eq.(\ref{globalAQ})
at $k=1\,.$ }. \medskip                                       

 All higher functions vanish $\acalk(0)=\agothk(0) =0$ in this limit. For
$k\geq 2\,,$ they oscillate in the IR region and form \cite{lmp99,tmp99}
an asymptotic sequence \`a l\'a Erd\'elyi.  \par
 The same properties remain valid for a higher--loop case. Explicit
expressions for \acalk and \agothk at the two--loop case can be written
down (see, Refs. \cite{mag00} and \cite{km01}) in terms of a special
Lambert function. They are illustrated below in Figs 1a and 1b.
   Note here that to relate Euclidean and Minkowskian functions, instead
of integral expressions (\ref{R-oper}) and (\ref{D-trans}) one can use
simpler relations, in terms of spectral functions
$\,\rho(\sigma)=\Im\acal(-\sigma) \,,$
\beglab{15}
{\cal A}_k(q^2;f)=\frac{1}{\pi}\int\limits_{0}^{\infty}\frac{d\sigma}
{\sigma+q^2}\:\rho_k(\sigma;f)\,;\quad\mathfrak{A}_k(s; f)=\frac{1}{\pi}
\int\limits^{\infty}_{s}\frac{d \sigma}{\sigma} \rho_k(\sigma; f)\,, \eeq
equivalent to expressions $\acalk(q^2)={\bf D}\left[\agothk\right]\,,$
and $\agothk(s) = {\bf R}[\acalk]\,.$ \par \smallskip

  Remarkably enough, the mechanism of liberation of unphysical
singularities is quite different. While in the space-like domain it involves
nonperturbative, power in $q^2$, structures, in the time-like region it is
based only upon resummation of the ``$\pi^2$ terms". Figuratively,
(non-perturbative~!) {\it analyticization} \cite{rapid96,prl97,tmp99} in
the $q^2$--channel can be treated as a quantitatively distorted reflection
(under $\,q^2\to s=\,-\,q^2\,$) of (perfectly perturbative)
$\,\pi^2$--resummation in the $s$--channel. This effect of ``distorting
mirror", first discussed in \cite{ms97} and \cite{mo98}, is clearly seen in
the figures 1a,b mentioned above. \par
  This means also that introduction of nonperturbative $\,1/q^2\,$
structures now has got \\ {\it another motivation, Eq.(\ref{acal}),
independent of the analyticization prescription.} \medskip

\subsection{Global APT}
In reality, a physical domain includes regions with various ``numbers of
active quarks", i.e., with diverse flavor numbers $f=3,4,5\,$ and 6. In
each of these regions, we deal with a different amount of quark quantum
fields, that is with distinct QFT models with corresponding Lagrangians.
To combine them into a joint picture, the procedure of the threshold
matching is in use. It establishes relations between renormalization
procedures for a model with different $f\,$ values. \par
  For example, in the \msbar scheme the matching relation has a simple form
\beglab{match}
\albars(q^2=M^2_f; f-1) =\albars(q^2=M^2_f; f)\,.\eeq
It defines a ``global effective coupling"
$$
\albars(q^2)=\albars(q^2;f)\,\quad\mbox{at}\,\quad M^2_{f-1}\leq q^2\leq
M^2_f\,,$$
continuous in the space-like region of positive $\,q^2\,$ values with
discontinuity of derivatives at matching points $q^2=M^2_f\,.$ To this
global \albars, there corresponds a discontinuous spectral density
\begin{equation}\label{discont}
 \rho_k(\sigma)=\rho_k(\sigma; 3) +\sum_{f\geq 4}^{}\theta(\sigma-M_f^2)
\left\{\rho_k(\sigma; f)-\rho_k(\sigma; f-1) \right\} \end{equation}
with $~\rho_k(\sigma; f)=\Im\,\albars^k (-\sigma, f)$ which yields
\cite{tow00,tmp01} via relations analogous to (\ref{15})
\begin{equation} \label{globalAQ}
{\cal A}_k(q^2)=\frac{1}{\pi}\int\limits_{0}^{\infty}\frac{d\sigma}
{\sigma+q^2}\:\rho_k(\sigma)\,;\quad\mathfrak{A}_k(s)=\frac{1}{\pi}
\int\limits^{\infty}_{s}\frac{d \sigma}{\sigma} \rho_k(\sigma)\,, \eeq
the smooth global Euclidean and spline--continuous global Minkowskian
expansion functions.\par

 \begin{figure}[th]
 \unitlength=1mm
   \begin{picture}(0,63)                                   %
   \put(0,1){
   \epsfig{file=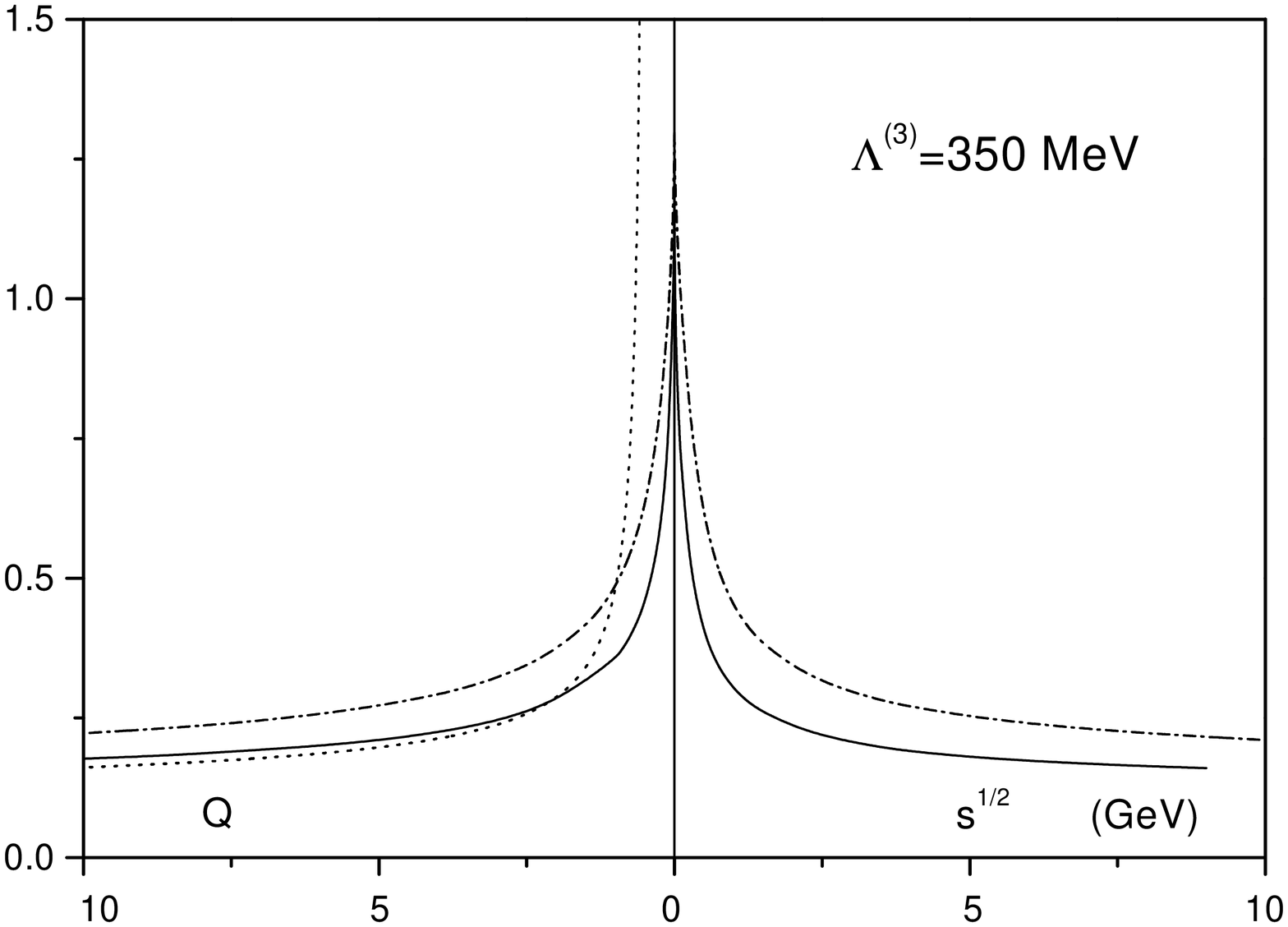,width=8.5cm}
   }
   \put(80,1){%
   \epsfig{file=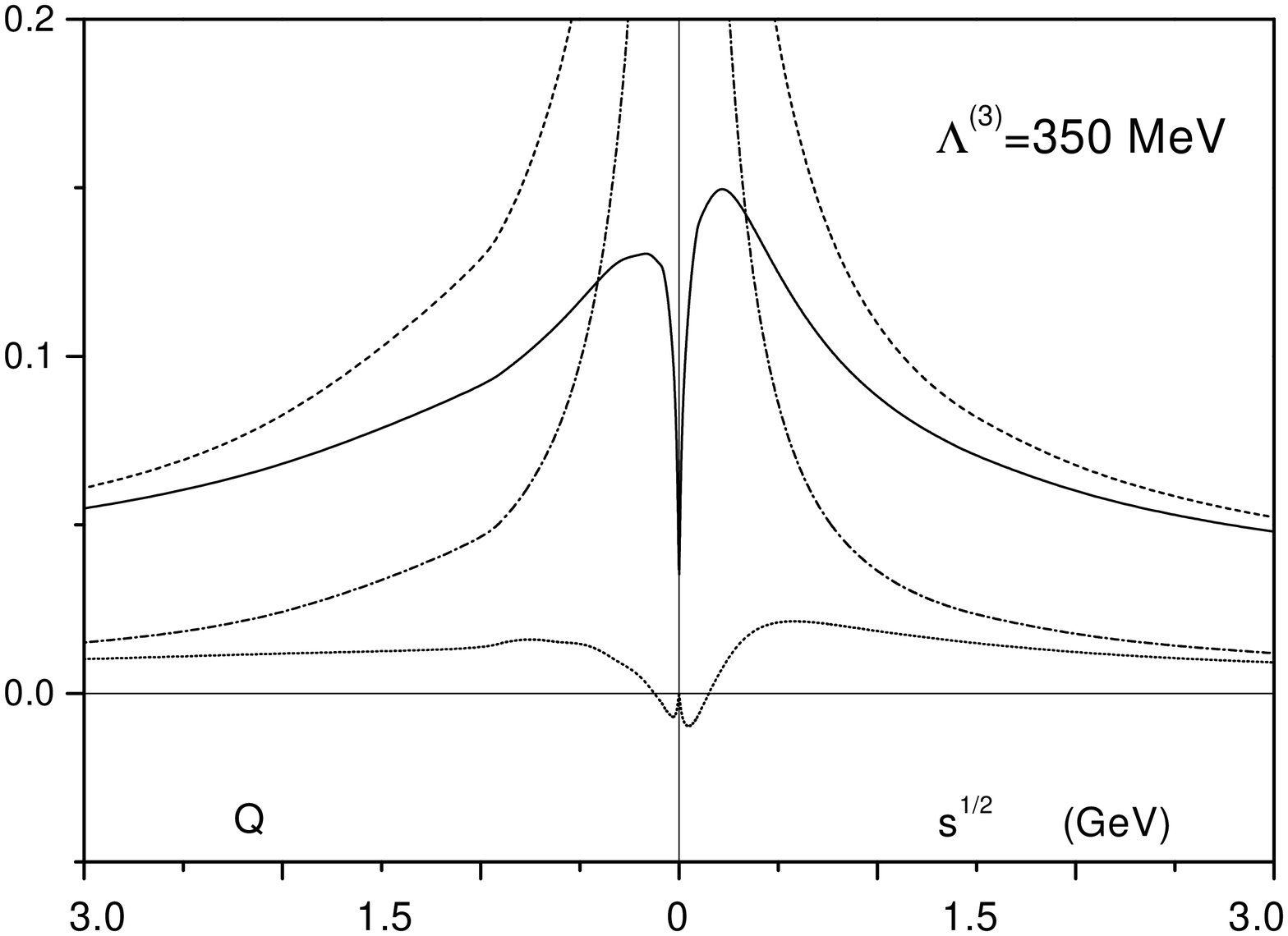,width=8.5cm}%
   }       %
 \put(14,53){\bf a}             %
 \put(94,53){\bf b}             %
  \put(43.8,53){$\bullet$}
  \put(27,53){\small $\alphan(0)=$}
  \put(48,53){\small \tildal(0)}
   \put(24,43){\small $\bar\alpha_s^{(2)}(q^2;3)$}
   \put(19,23){\small  $\alpha_{\rm{an}}^{(1)}(q^2;3)$}
   \put(33,14){\small $\alpha_{\rm an}^{(2,3)}$}
   \put(51,23){\small $\tilde{\alpha}^{(1)}(s;3)$}
   \put(46,13){\small $\tildal^{(2,3)}$}
    \put(110,50){\small $\alpha_{\rm an}^2$}
    \put(112,39){\small\bf ${\cal A}_2$}
    \put(101,26){\small $\alpha_{\rm an}^3$}
    \put(116,24){\small\bf ${\cal A}_3$}
    \put(138,39){\small $\tilde{\alpha}^2$}
    \put(133,33){\small ${\mathfrak A}_2$}
    \put(145,24){\small  $\tilde{\alpha}^3$}
    \put(127,25){\small ${\mathfrak A}_3$}
  \end{picture}

\centerline{    \parbox{15.5cm}{
\caption{\sl\footnotesize {\bf a} -- Space-like and time-like global
analytic couplings in a few $GeV$ domain with $f=3$ and $\Lambda^{(3)}=
350 \:\MeV$; {\bf b} -- ``Distorted mirror symmetry" for global expansion
functions. All the curves in {\bf 1b} correspond to exact two--loop
solutions expressed in terms of Lambert function.} \label{fig1} }}
 \end{figure}

Here, in Fig.1a, by the dotted line we give a usual two-loop effective QCD
coupling \asq with a singularity at $\,q^2=\Lambda^2.$ Meanwhile, the
dash--dotted curves represent the one-loop APT expressions (\ref{tildal1})
and (\ref{4}). The solid APT curves are based on the exact two-loop solutions
of RG equations and approximate three--loop solutions in the \msbar scheme.
Their remarkable coincidence (within the 2--4 per cent) demonstrates reduced
sensitivity of the APT approach (see, also Refs. \cite{prl97,ss98pl,alex01})
with respect to higher--loop effects in the whole Euclidean and Minkowskian
regions from IR to UV limits. Fig.1b shows higher two--loop functions
 in comparison with \alphan and \tildal powers. \par

  Generally, functions \agothk and \acalk differ from the local ones
with a fixed $f\,$ value. Minkow\-ski\-an global functions \agothk can be
presented via $\agothk(s, f)\,$  by relations
\begin{equation}\label{al-s-f}
\tildal(s)\,=\tildal(s; f)\,+c(f)\,;\quad \agoth_2(s)=\agoth_2(s; f)+
\:{\mathfrak c}_2(f) \quad
\mbox{at}\quad \quad M^2_f\leq s \leq M^2_{f+1}\,  \eeq
with {\it shift constants} $c(f), \:{\mathfrak c}_2(f)$ representable via
integrals over $\rho_k(\sigma; f+n)\:,\,n\geq 1\,$ with additional
reservations, like $\,c(6)=0\,,$ related to the asymptotic freedom
condition.  \par
Numerical estimate performed in Ref.\cite{tmp01} (see also Table 6 in
Ref.\cite{mag00}) for traditional values of the QCD scale parameter
$\,\Lambda_{3}\sim 300-400$ \MeV \
 $$
~c(3)\sim 0.02\,,~c(4)\simeq 3.10^{-3}\:,\,c(5)\simeq 3.10^{-4}\:;\,
\quad \:{\mathfrak c}_2(f)\simeq 3\,\alpha(M_f^2)\,c\,(f) \,$$
reveals that these constants are essential in the $f=3,4\,$ region at a few
per cent level for $\,\tildal\,$ and at ca 10\% level for $\agoth_2\,$.\par
  Meanwhile, global Eucledean functions $\acalk(q^2)\,$ cannot be related
to the local ones $\acalk(q^2, f)\,$ by simple relations. Nevertheless,
numerical calculation shows \cite{mag00,km01} that in the $f=3\,$ region
one has approximately
\begin{equation}\label{cal-s-f}
\alphan(q^2)\,=\alphan(q^2; 3)\,+c(3)\,;\quad \acal_2(q^2)=\acal_2(q^2); 3)+
\:{\mathfrak c}_2(3) \quad
\mbox{at}\quad M^2_3\leq s \leq M^2_{4}\,.  \eeq

\section{\sf The APT applications\label{s3}}                            
\subsection{General comments\label{s3.1}}
In what follows, we abstract ourselves of recent successive use of the
Analytic approach to hadronic formfactors\cite{nicos} and concentrate on
the QCD applications of APT.

To illustrate a quantitative difference between global APT scheme and
common practice of data analysis in perturbative QCD, consider a few
examples. \par

 In the usual treatment --- see, e.g., Ref. \cite{pdg00} --- the (QCD
perturbative part of)  Minkow\-ski\-an observable, like $e^+e^-$
annihilation or $Z_0$ decay cross--section ratio, is presented in the form
\begin{equation}\label{Rtrad}
R(s)=R_0 \left(1+r(s)\right)\,; \quad r_{PT}(s)=c_1\,\albars(s)+c_2\,
\albars^2(s)+ c_3\,\albars^3(s)+ \dots \,\:. \end{equation}

 Here, the coefficients $\, c_1\,,~\,c_2\, $ and $~c_3\,$ are
not diminishing numerically --- see Table 1. A rather big negative $c_3$
value comes mainly from the $\,-c_1\pi^2\beta^2_0/3\,\,$ term. In the APT,
we have instead
 \begin{equation}\label{rapt}
r_{APT}(s)=d_1\tildal(s)+d_2\,{\mathfrak A}_2(s) +d_3\:{\mathfrak A}_3(s)\:
+ \dots \eeq
with reasonably decreasing Feynman coefficients $\,d_{1,2}=c_{1,2}\,$ and
$~d_3=c_3+c_1\pi^2\beta^2_0/3\,,$ the mentioned $\pi^2$ term
of $c_3$ being ``swallowed" by $\,\tildal(s)\,.$ \medskip

In the Euclidean channel, instead of power expansion similar to
(\ref{Rtrad}), we typically have
 \begin{equation}\label{dapt}
d_{APT}(q^2)=d_1\alpha_{\rm an}(q^2) +d_2\,{\cal A}_2(q^2)+
 d_3\,{\cal A}_3(q^2)\:+ \dots  \end{equation}
with the same coefficients $d_k\,$ extracted from Feynman diagrams. Here,
the modification is related to nonperturbative, power in $q^2$, structures
like in (\ref{4}).

\begin{center}
 {\sf Table 2 : {\small Relative contributions (in \%)
of 1-- , 2-- and 3--loop terms to observables}}
\label{tab2}\smallskip

\begin{tabular}{|l|l|l||c|c|c||r|r|r|}  \hline
{\slshape \phantom{aa}Process} & {\slshape $q$ or $\sqrt{s}$}&
{\slshape f} & \multicolumn{3}{|c|}{\slshape PT} &
\multicolumn{3}{r|}{\slshape APT\phantom{aaaa}} \\ \hline\hline
GLS sum rule & 1.73\, \GeV & 4 & 65 & 24 & 11 & 75 & 21 & \bf{4}\\ \hline
Bjorken. s.r.& 1.73\, \GeV & 3 & 55 & 26 & 19  &  80 & 19 & \bf{1}  \\
\hline\hline Incl. $\tau$-decay & 0 - 2 \GeV & 3 & 55 & 29 & 16 & 88 & 11 &
\bf{1}\\ \hline
$e^+e^-\to$ hadr. & 10 \GeV & 4 & 96 &  8 & -4  &  92 &  7 & \bf{.5} \\
  \hline
$Z_o \to$ hadr.  & 89 \GeV & 5 & 98.6 & 3.7 & -2.3 & 96.9 & 3.5 & -\bf{.4} \\
\hline \end{tabular} \end{center} \medskip

In Table 2, we give values of the relative contribution of the first, second,
and third terms of the r.h.s. in (\ref{Rtrad}),(\ref{rapt}) and (\ref{dapt})
for the Gross--Llywellin-Smith \cite{mss99} and Bjorken \cite{mss98}
sum rules, $\tau$ -- decay in the vector channel \cite{mss01}, as well as
for $e^+e^-$ and $Z_0$ inclusive cross-sections. As it follows from this
Table, in the APT case, the three--loop (last) term is very small, and
being compared with data errors, numerically inessential. This means that,
in practice, \medskip                                              

\centerline{\sf\normalsize one can use the APT expansions (\ref{rapt}) and
(\ref{dapt}) without the last term.}  \smallskip

\subsection{Semi--quantitative estimate \label{3.2}}                
  This conclusion can be valuable for the case when the three--loop
contribution, i.e., $d_3$ is unknown. Here, people use the so-called NLLA
approximation, that is common practice in the $f=5\,$ region. For the
Minkowskian observable, e.g., in the event--shape (see, e.g., Ref.
\cite{delphi}) analysis there corresponds the two-term expression
 \beq\label{two-term}
r(s) = c_1 \as(s) +c_2 \as^2(s)\,. \eeq

 On the basis of the numerical estimates of Table 1, in such a case, we
recommend instead {\it to use the two-term APT representation}
 \begin{equation}\label{r2apt}
 r^{(2)}_{APT}(s)=d_1\tildal(s)+d_2\,{\mathfrak
A}_2(s) \eeq
which, at $L^2\gg \pi^2$, is equivalent to the three-term expression
\beq\label{3Delta}
r_3^{\Delta}(s)=d_1\left\{\albars-\frac{\pi^2\beta_0^2}{3}\albars^3\right\}
+d_2\albars^2 =c_1\,\albars+ c_2\albars^2- \underline{{\bf\delta_3\,
\albars^3}}\,,\eeq
i.e., to take into account the known predominant \pisq part of the next
coefficient $c_3\,$. As it follows from the comparison of the last
expression with the previous, two--term one (\ref{two-term}), the \albars
numerical value extracted from eq.(\ref{3Delta}), for the same measured
value $r_{obs}$, will differ mainly by a positive quantity (e.g., in the
$f=5\,$ region with $\,\albars \simeq 0.12 \div 0.15)$
\beglab{del3}
(\triangle\albars)_3=\left.\frac{\pi\delta_3\,\albars^3}{1+2\pi d_2
\albars}\right|_{20\div 100 \GeV}^{f=5}=\frac{1.225\,\albars^3}
{1+0.90\,\albars}\simeq0.002\div0.003\, \eeq
that turns out to be numerically important. \par \smallskip

 Moreover, in the $f=4\,$ region, where the three-loop (NNLLA) approximation
is commonly used in the data analysis, the \pisq term $\delta_4\,$ of the
next order turns out also to be essential. Hence, we propose there, instead
of (\ref{Rtrad}), {\it to use the APT three--term expression}
 \begin{equation}\label{r3apt} r^{(3)}_{APT}(s)=d_1\tildal(s)+
d_2\,{\mathfrak A}_2(s)+d_3\,{\mathfrak A}_3(s) \eeq
approximately equivalent to the four-term one
\beq\label{4term}
r_4^{\Delta}(s)= d_1\,\albars+ d_2\,\albars^2 +c_3\,\albars^3-
\underline{\mathbf{\delta_4\,\albars^4}}\,; \quad c_3=d_3-\delta_3\,\eeq
 or to {\small
$$
r_4^{\Delta}(s)= d_1\left\{\albars-\frac{\pi^2\beta_0^2}{3}\albars^3
-b_1\frac{5}{6}\,\pi^2\beta_0^2\,\albars^4\right\} +d_2\left\{\albars^2
-\pi^2\beta_0^2\,\albars^4\right\}+ d_3\,\albars^3 $$}
with $\delta_3$ and $\delta_4$ defined\cite{rad82,kat95} in eq.(\ref{deltas}).
\par                                                               

 The three-- and two--term structures in braces are related to specific
expansion functions $\tildal=\agoth_1$ and $\agoth_2$ defined above
(\ref{globalAQ}) and entering into the non-power expansion (\ref{r3apt}).
\medskip

  To estimate roughly the numerical effect of using this last modified
expression (\ref{4term}), we take the $e^+e^-$ inclusive annihilation. For
$\sqrt{s}\simeq 3\div5\:\GeV\,$ with $\albars\simeq 0.28\div 0.22\,$ one has
$$
\left.(\triangle\albars)_4=\frac{\pi\delta_4\,\albars^4}{1+2\pi d_2
\albars}\,\right|_{3\div 5 \GeV}^{f=4} =\frac{1.07\,\albars^4}{1+0.974\,
\albars}\simeq 0.005 \div 0.002 $$
--- an important effect on the level of ca $1 \div 2 \% \,.$

 Moreover, the $(\triangle\albars)_4\,$ correction turns out to be
noticeable even in the lower part of the $f=5\,$ region! Indeed, to
$\sqrt{s}\simeq 10\div 40 \: \GeV\,$ with $\albars \simeq 0.20 \div
0.15\,$ there corresponds
$$
\left.(\triangle\albars)_4\right|_{10\div 40\:\GeV}^{f=5} \simeq 0.71 \,
\albars^4\simeq (1.1\div 0.3) \cdot 10^{-3} \quad (\lesssim 0.5\%)\,.$$

\subsection{Important warning\label{ss3.3}}
It is essential to note that approximate expressions eqs.(\ref{3Delta})
and (\ref{4term}) are equivalent to the exact ones (\ref{r2apt}) and
(\ref{r3apt}) only in the region $\,L=\ln\left(s/\Lambda^2\right)\gg\pi\,$
as it is shown on Fig. 2.
\begin{figure}[ht]\label{fig-2} \unitlength=1mm
\begin{picture}(155,100)   \put(-3,-9){%
\centerline{\epsfig{file=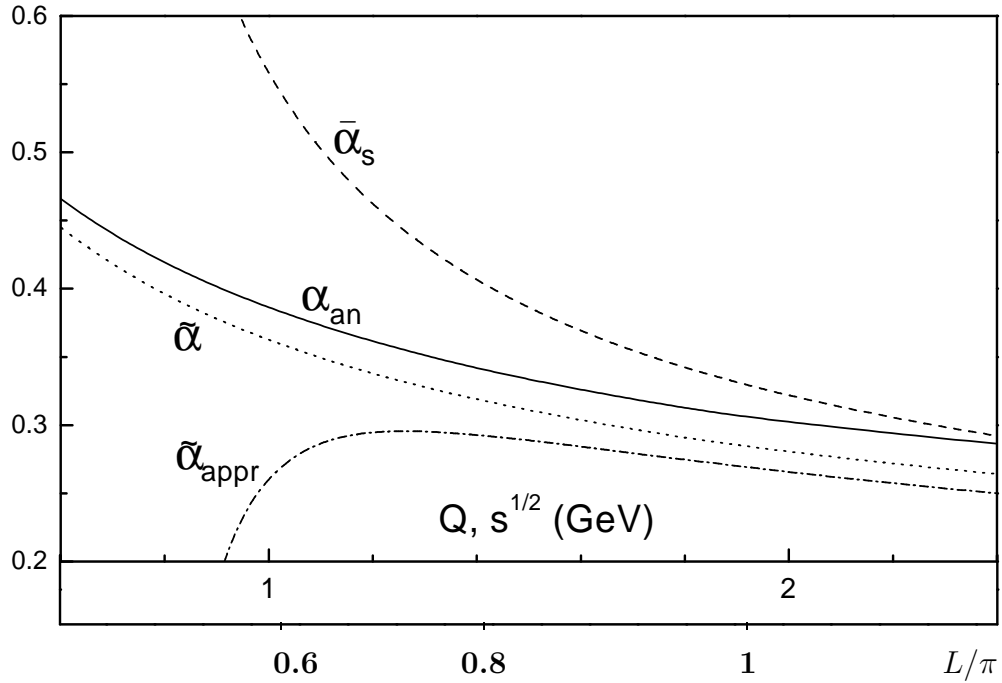,width=155mm,height=110mm,angle=0}}%
}
\put(50,7.8){\line(0,1){1}}
 \put(49,2){\bf 0.6}%
\put(77,7.8){\line(0,1){1}}
 \put(74,2){\bf 0.8}%
\put(111,2){\bf 1}             %
 \put(112,7.8){\line(0,1){1}}                              %
 \put(138,2){\bf $L/\pi$}%
\end{picture} %
\centerline{ \parbox{15cm}{ \caption{\footnotesize Comparison of common
QCD coupling \albars with the APT global ones (\tildal, \alphan) in the
$\,q,\sqrt{s}< 3\,\GeV\,$ region at $\,\Lambda_3 =400 \ \MeV\,$. By
dash-dotted line we give an approximate Minkowskian coupling (\ref{app2}).
All the curves are taken (see Tables 1,5 and 6 in Ref.\cite{km01})
for the 3-loop global case.}} }
\end{figure}

 One can see that the curve for approximate Minkowskian coupling
\beglab{app2} \tildal_{appr}(s)=\albars(s)-(\pi^2\beta_0^2/3)
\albars^3\,,\eeq that precisely corresponds to the popular
approximation (\ref{Rtrad}) (and gives rise to the $\pi^2\,$term
in the $\as^3$ coefficient) has a rather peculiar behavior. In the
region $\,L > \pi\,$ it goes rather close to the curve for
\tildal. For instance, at $\,L \simeq \pi\,$ the relative
error of approximation is about 5 per cent. On the other hand, below
$\,L\simeq 0.8\,\pi\,$ (i.e., $W\simeq 1.0-1.4\,\GeV\,$) the
distance $\,\tildal-\tildal_{\rm appr}\,$ between curves (error of
approximation) increases and at $\,L\simeq 0.7\,\pi\,$ it blows up
(better to say ``comes down").\par

In particular, at $\,s \leq 2\,\GeV^2\,$ it is rather desorienting to
refer to $\,\albars(s)\,$ and it is erroneous to use
$\,\tildal_{appr}(s)\,$ and common expansion (\ref{Rtrad}).\par
This means that
\medskip

\centerline{\sf below $\,s=2\,\GeV^2\,$ it is nonadequate to use common
$\albars(s)\,$ and power expansion eq.(\ref{Rtrad}).} \par \smallskip

In other words, we claim that below $\,s=2\, \GeV^2\,$ it is an intricate
business to analyze data in terms of the ``old good" (but singular) \as
\footnote{In particular, this relates to analysis of $\tau$ decay. In this
connection we would like to attract attention to the important paper
\cite{mss01} that treats the $\tau$ decay within the APT approach (with
effective mass of light quarks and the threshold resummation factor) and
results in $\Lambda^{(3)}= 420 \ \MeV\,$ that corresponds to
$\,\alphan(M^2_{\tau})= 0.32\,$ or $\tildal(M^2_{\tau})= 0.30\,.$
At the same time, attempts to interpret results of APT for $\tau$ decay in
terms of $\as,$ like, e.g., in Ref.\cite{nonsholom}, needs some special
precaution --- see next footnote. A more detailed comment on the $\,\tau\,$
decay theoretical analysis will be published elsewhere.}.
Here, approximate relation (\ref{app2}) does not work as it is illustrated
in Fig.2.\par
 In this, low--energy Minkowskian/Euclidean region data have to be analyzed
in terms of nonpower expansion (\ref{rapt})/(\ref{dapt}) and extracted
parameter should be $\alphan(s)\,, \tildal(q^2)\,$ or $\Lambda^{(3)}\,.$
In Table 3 we give few numerical examples for the chain
$$
\alphant\leftrightarrow\tildalt\leftrightarrow\Lambda^{(3)}\to\Lambda^{(5)}
\leftrightarrow\albarsZ$$
that allows to study the QCD theoretical compatibility of LE data with
the HE ones in the APT analysis.
\begin{center}
 {\sf Table 3 : {\small Numerical chain related LE with HE regions}}
\smallskip

\begin{tabular}{|l|l|l||r|r|}  \hline
$\tildalt$&$\alphant$&\phantom{aaa}$\Lambda^{(3)}$&$\Lambda^{(5)}$
\phantom{aa}& $\albarsZ$ \\ \hline \hline
0.309    &   0.332  & 450 \MeV      & 303 \MeV       & 0.125    \\ \hline
0.292    &   0.314  & 400 \MeV      & 260 \MeV       & 0.121   \\ \hline
0.278    &   0.299  & 350 \MeV      & 218 \MeV       & 0.119    \\ \hline
0.266    &   0.286  & 300 \MeV      & 180 \MeV       & 0.116    \\ \hline
\end{tabular} \end{center} \medskip
Here, the main element of correlation is the chain
$\Lambda^{(3)}\leftrightarrow\Lambda^{(3)}\leftrightarrow\Lambda^{(5)}$
that follows from the matching condition (\ref{match})
\footnote{Generally, it is possible to use correspondence between
\alphan, \tildal and \as as expressed by relations (\ref{altrans}).
However, the use of $\as^{\rm \msbar}(\mu^2)\,$ at $\,\mu \lesssim 1\
GeV\,$ as a QCD parameter could be misleading due to vicinity to
singularity. For example, at $\Lambda^{(3)}= 400 \ \MeV \,$ one has
$\,\as(M^2_{\tau})\simeq 0.35\,$ and $\,\as(1\ \GeV^2)\simeq 0.55\,$
to be compared with $\,\alphan(M^2_{\tau})\simeq 0.31\,$ and
$\,\alphan(1\ \GeV^2)\simeq 0.40\,.$}.

\section{\sf  Quantitative illustration \label{s4}}                
 Consider now a few cases in the $f=5\,$ region. \par
\medskip
\underline{$\Upsilon$ decay.} According to the Particle Data Group
(PDG) overview (see their Fig.9.1 on page 88 of Ref.\cite{pdg00}),
this is (with $\as(M^2_{\Upsilon})\simeq 0.170$ and $\asmz=0.114$)
one of the most ``annoying" points of their summary of \asmz
values. It is also singled out theoretically. The expression for
the ratio of decay widths starts with the cubic term\footnote{See,
e.g., eq.(9.16) in Ref.\cite{pdg00}.}
$$
R(\Upsilon)=R_0\,\as^3(\xi M_{\Upsilon}^2)(1-e_1\,\as)\quad\mbox{with}
\quad \xi\lesssim 0.5\quad\mbox{and}\quad c_1(\xi)\,\simeq 1\,\,.$$
Due to this, the $\pi^2$ corrections\footnote{First proposal of taking
into account this effect in the $\Upsilon$ decay was discussed
\cite{kras82}
more than a quarter of century ago. Nevertheless, in current practice
it is completely forgotten.} corresponding to the APT expression
\beglab{31}
R_{\rm APT}(\Upsilon)=R_0\,\agoth_3(\xi M_{\Upsilon}^2)(1-e_1\,\agoth_4)\eeq
are rather big
$\agoth_3\simeq \as^3\left(1-2(\pi\beta_0)^2\as^2 \right),\,\,
\agoth_4\simeq\as^4\left[1-(10/3)(\pi\beta_0)^2\as^2\right]\,$
in the region with $\pi^2\beta_0^2(5)=3.57$ and
$\as(\xi M_{\Upsilon}^2)\simeq 0.2\,.$
As a crude estimate (taken from $\as^3 \to \agoth_3$ only),
$$\Delta\as(M_{\Upsilon}^2)=\frac{2}{3}\,(\pi\beta_0)^2\,
\as^3(M^2_{\Upsilon}) \simeq 0.0123 \,,$$ which corresponds to
\beglab{del-ups-asz} \Delta\albars(M^2_Z)=0.006 \quad\mbox{with
resulting}\quad\albars(M^2_Z)= 0.120\,.\eeq
 One should note here, that this estimate is rather crude and gives
only indication of the order of magnitude.

\underline{The NNLO case.} Now, let us turn to a few cases analyzed by the
three-term expansion formula (\ref{rnew}). For the first example, take
\underline{\sf $e^+e^-$ hadron annihilation} at $\sqrt{s}=42\,\GeV$ and
$11\,\GeV\,.$

A common form (see, e.g., Eq.(15) in Ref.\cite{beth00}) of theoretical
presentation of the QCD correction in our normalization looks like
$$
r_{e^+e^-}(\sqrt{s})=0.318\albars(s)+0.143\,\albars^2 -0.413\,\albars^3\,.$$
 In the standard PT analysis, one has (see, e.g., Table 3)
$\albars(42^2)=
0.144\,$ that corresponds to $r_{e^+e^-}(42)\simeq 0.0476\,.$  Along with
the APT prescription, one should use
\beq\label{r-annih}
r_{e^+e^-}(\sqrt{s})=0.318\,\tildal(s)+0.143\,\agoth_2(s)-0.023\,\agoth_3(s)\,,\eeq
which yields $\tildal(42^2)=0.142\,\,\to \,\as(42^2)=0.145$ and
$\asmz=0.127\,$
to be compared with $\asmz=0.126\,$ under a usual analysis. \par \smallskip

 Quite analogously, with $\albars(11^2)=0.200\,$ and $r_{e^+e^-}(11)\simeq
0.0661\,$ we obtain via (\ref{r-annih}) $\tildal(11^2)=0.190\,$ that
corresponds to $\asmz=0.129\,$ instead of 0.130. \smallskip

  For the next example, take the \underline{\sf $\/Z_0$ inclusive decay}.
The observed ratio
$$
R_Z=\Gamma(Z_0\to hadrons)/\Gamma(Z_0\to leptons)=20.783 \pm.029\,$$
 can be written down as follows:
$R_Z=R_0\left(1+r_Z(M_Z^2)\right)\,$ with $R_0=19.93\,.$ A common form
(see, e.g., Eq.(15) in Ref.\cite{beth00}) of presenting the QCD
correction $r_Z$ looks like
$$
r_Z(M_Z)=0.3326\albars + 0.0952\,\albars^2 -0.483\,\albars^3\,.$$

  To $\left[r_Z\right]_{obs}= 0.04184\:$ there corresponds
$\:\asmz=0.124\,$ with $\,\Lambda_{\msbar}^{(5)}=292 \,\MeV\,.$
In the APT case, from
\beq\label{r-pi}
r_Z^{\rm APT}(M_Z)=0.3326\,\tildal(M_Z^2)+0.0952\,\agoth_2(M_Z^2)-0.094\,
   \agoth_3(M_Z^2)\eeq
we obtain $\,\tildal(M_Z^2)= 0.122\,$ and $\:\asmz=0.124\,.$
Note that here the three-term approximation (\ref{r-apt}) gives the
same relation between the $\:~\asmz\,$ and $\:\tildal(M^2_Z)\,$ values.

 Nevertheless, in  accordance with our preliminary estimate for
the $\,(\triangle\albars)_4\,$ role, even the so-called NNLO theory needs
some  $\pi^2$ correction in the $W=\sqrt{s}\lesssim 50\:\GeV$ region.\par
\medskip

\underline{The NLO case.} Now, turn to those experiments in the HE $(f=5)$
Minkowskian region (mainly with a shape analysis) that usually are confronted
with the two-term expression (\ref{two-term}). As it has been shown above
(\ref{del3}), the main theoretical error here can be expressed in
\bigskip

\begin{minipage}[bt]{150mm}
\begin{center}
{\sf\large Table 4 : \label{tab4}} 
{\sf The APT revised\footnote{``j \ \& \ sh" = jets and shapes;
Figures in brackets in the last column give the \\ dif\-ference
$\Delta\asmz$ between common and APT values.}
part ($f=5$) of Bethke's \cite{beth00} Table 6 } \medskip

\begin{tabular}{|c|c|c||c|c||c|c|}  \hline \smallskip

&$\sqrt{s}$&loops&\albars(s)&\asmzs&\albars(s)&\asmzs\\ Process&\GeV& No&
ref.[2] &ref.[2]& APT &APT  \\ \hline \hline
$\Upsilon$-decay
\footnote{Taken from Ref.\cite{pdg00}.}
                       &9.5 &2      &.170&.114&.182&.120 (+6)     \\
$e^+e^-[\sigma_{had}]$ &10.5&{\bf 3}&.200&.130&.198&.129({\bf -1})\\
$e^+e^-[j\, \&\, sh]$  &22.0&2 &.161&.124&.166&.127(+3)  \\
$e^+e^-[j\, \&\, sh]$  &35.0&2 &.145&.123&.149&.126(+3)  \\
$e^+e^-[\sigma_{had}]$ &42.4&{\bf 3}&.144&.126&.145&.127(+{\bf1})\\
$e^+e^-[j\, \& sh]$    &44.0&2 &.139&.123&.142&.126(+3) \\
$e^+e^-[j\, \& sh]$    &58  &2 &.132&.123&.135&.125(+2) \\
{\bf $Z_0\to$ had.}    &91.2&{\bf 3}&.124&.124&.124&.124({\bf 0}) \\
$e^+e^-[j\,\&\,sh]$    &91.2&2      &.121&.121&.123&.123(+2) \\
-"-                    &....& 2     & ...& ...& ...& ...(+2)\\
$e^+e^-[j\,\&\,sh]$    &189 &2      &.110&.123&.112&.125(+2)\\ \hline
\end{tabular}\end{center}
Averaged $<\asmzs>_{f=5}$ values \hspace{24mm}\/$0.121;$
\hspace{18mm} $0.124\,$
\end{minipage} \smallskip

\noindent the form
\beglab{das5}
\left.(\triangle\albars(s)\right|_{20\div 100 \GeV}^{f=5}\simeq
1.225\,\albars^3(s)\simeq0.002\div0.003\,.\eeq
An adequate expression for the equivalent shift of the $\asmz$ value is
\beglab{dasmz}
[\triangle\asmz]_3 =1.225\albars(s)\asmz^2\,.\eeq

We give the results of our approximate APT calculations, mainly by
Eqs.(\ref{das5}) and (\ref{dasmz}), in the form of Table 4 and Figure 3. In
the last column of Table 4 in brackets, we indicate the difference between
the APT and usual analysis. The results of the three--loop analysis are
marked by bold figures.  Dots in the lower part of the Table correspond to
shape--events data for energies $\,W= 133, 161, 172 \,$ and $\,183 \,\GeV\,$
with the same positive shift 0.002 for the the extracted \albars values.\par

   \begin{figure}[hbt]\label{fig-3}
\unitlength=1mm
\begin{picture}(0,95)
  \put(18,10){%
   \epsfig{file=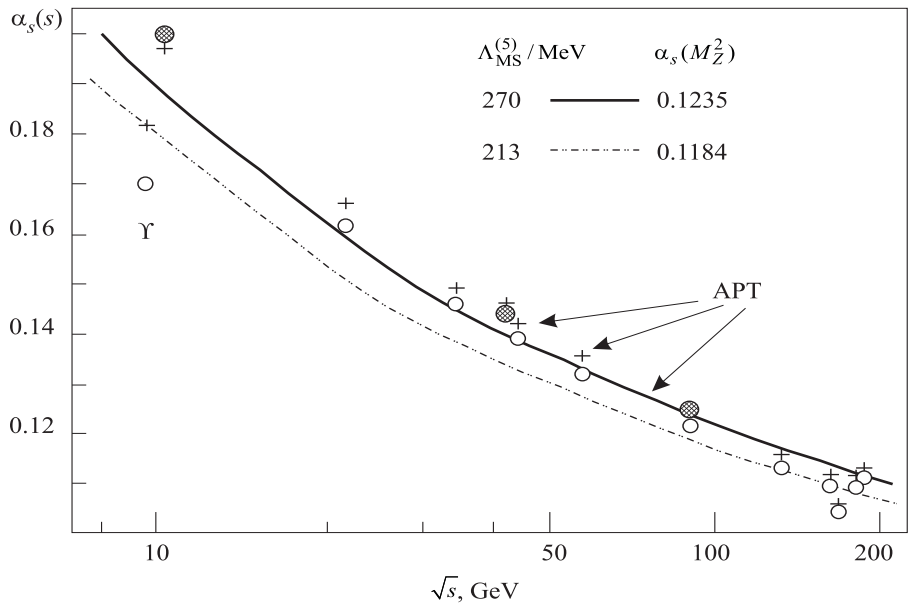,width=120mm,height=80mm}%
}
\end{picture}
\centerline{ \parbox{15cm}{ \caption{\footnotesize
 The APT analysis for \albars in the $f=5$ time--like region. Crosses (+)
differ from circles by the $\pi^2$ correction (\ref{das5}). Solid APT curve
relates to $\Lambda_{\rm MS}^{(5)}=270 \,\MeV$ and $~\asmz=0.124\,.$
By dot-and-dash curve, we give the standard \albars (at $\Lambda^{(5)}=213
\,\MeV$ and $~\asmz=0.118$) taken from Fig.10 of paper \cite{beth00}.}}}
\end{figure}

 In Fig.3, by open and hatched circles we give two--
and three--loops data from Fig.10 of paper \cite{beth00}. The only exclusion
is the $\Upsilon$ decay taken from Table X of the same paper. By crosses, we
marked the ``APT values" calculated approximately by Eq.(\ref{das5}).

For clearness of the \pisq effect, we skipped the error bars. They are
the same as in the mentioned Bethke's figure and we used them for
calculating $\chi^2\,.$

 Let us note that our average 0.121 over events from Table 6 of Bethke's
review \cite{beth00} nicely correlates with recent data of the same author
(see Summary of Ref.\cite{beth-fest}). The best $\chi^2$ fit yields
$\asmz_{[2]}=0.1214$ and\footnote{This value, corresponding to
$\Lambda^{(5)}= 290 \,\MeV\,,$ is supported by recent analysis \cite{mss01}
of $\tau$ decay that gives $\Lambda^{(3)}= 420\: \MeV\,;$ compare with
Table 3.}
$$\asmz_{APT}=0.1235\,.$$
 This new $\chi^2_{\rm APT}$ is smaller $\chi^2_{APT}/\chi^2_{PT}\simeq
0.73$ than the usual one. This illustrates the effectiveness of the APT
procedure in the region far enough from the ghost singularity.\par

\section{\sf Conclusion \label{s5}}
 It is a common standpoint that in QCD it is legitimate to use the power
in \as expansion for observables in the low energy (low momentum transfer)
region. At the same time, there exist rather general (and old
\cite{Dyson52}) arguments in favor of nonanalyticity of the $S$ matrix
elements at the origin \cite{lmp76} of the complex plane of the expansion
parameter $\alpha\,$ variable.  This, in turn, implies that common
perturbation expansion has no domain of convergence. Technically, this
corresponds to the factorial growth ($\sim n!\,$) of expansion coefficients
(like $d_n$ or $r_n\,$) at large $\,n\,$\cite{lip77,fdp80}. In QCD, with
its ``not small enough" \as values in the region below $10\ \GeV\:$ it is
a popular belief that one does face an asymptotic nature of perturbation
expansion by observing approximate equality of relative contributions of
the second ($\as^2$) and the third ($\as^3$) terms into observable, like
in all PT columns of Table 2. \par
   Our first  qualitative result consists in observation that convergence
properties of the APT expansions drastically differ from the usual PT ones.
\par
  The evidently better practical convergence of the APT series for the
Euclidean observable, as it has been demonstrated in the right part of Table
2, probably means that essential singularity at $\,\as=0\,$ is adequately
taken into account by new expansion functions $\,\acalk(q^2)\,.$ On the
other hand, in the time--like region the improved approximation property
of the APT expansion over $\,\agothk(s)\,$ has a bit different nature,
being related, in our opinion, to the non-uniform convergence of the
standard PT expansion for Minkowskian observables. In any case, from a
practical point of view: \par\smallskip

\noindent {\bf 1.~}{\sf In the APT, one can use the nonpower
expansions (\ref{rapt}) and (\ref{dapt}) without the last term.}\par
\medskip

 The next point, discussed in Section \ref{ss3.3}, refers to a more
specific issue connected with current practice of the Minkowskian observable
analysis in the low--energy $\,(s\lesssim 3\, \GeV^2)\,$ region (like, e.g.,
inclusive $\,\tau\,$ decay). As it has been shown -- see Fig. 2 ---
\par\medskip

\noindent {\bf 2.} {\sf Below 2 $\GeV^2$ it is impossible to use the
common power expansion (\ref{rnew}) for a time--like observable.}\bigskip

 Second group of our results is of a quantitative nature:\par\medskip

\noindent {\bf 3.} Effective positive shift $\,\Delta\albars\simeq +0.002\,$
in the upper  half ($\geq \, 50 \,\GeV$) of the $f=5\,$ region for all
time-like events that have been analyzed up to now in the NLO mode.\par
\par\medskip   

\noindent{\bf 4.} Effective shift $\Delta\albars \gtrsim +0.003\,$ in the
lower half ($10 \div 50 \, \GeV$) of the $f=5\,$ region for all time-like
events that have been analyzed in the NLO mode.\par
\medskip

\noindent {\bf 5.} The new value
\beq\label{124}
 \asmz=0.124\,, \eeq
obtained by averaging new APT results over the $f=5$ region. \par\medskip

  The quantitative results are based on the new APT nonpower expansion
(\ref{r-apt}) and plausible hypothesis on the $\pi^2\,$--term prevalence
in common expansion coefficients for observables in the Minkowskian domain.
The hypothesis has some preliminary support --- see Table 1 --- but needs
to be checked in more detail. \par
 Nevertheless, our result (\ref{124}) being taken as granted raises two
physical questions: \par\smallskip

 -- The issue of self-consistency of QCD invariant coupling behavior
between the ``medium $(f=3,4)$" and ``high  $(f=5,6)$" regions. \par
    Here, the more detailed APT analysis of data on DIS, heavy quarkonium
decays and some other processes are in order. As it has been mentioned
above, fresh APT analysis of $\,\tau$--decay \cite{mss01} seems to support
such a correlation with $\,\Lambda_3\sim 400\div 450\, \MeV$ and
$\,\Lambda_5\sim 290\, \MeV.$ \medskip

 -- The new ``enlarged value" (\ref{124}) can influence various physical
speculations in the very HE region, in particular on superpartner masses
in MSSM GUT constructions  -- compare, e.g., with recent
attempts \cite{kazgut1} in this direction. \bigskip

{\bf\large Acknowledgements} \smallskip

 The author is indebted to  A.P. Bakulev, D.Yu.~Bardin, Yu.L. Dokshitzer,
 F. Jegerlehner, B.~Kniehl, R.~K\"{o}gerler, N.V.~Krasnikov, B.A.~Magradze,
S.V. Mikhailov, I.L.~Solovtsov, O.P.~Solovtsova and N. Stefanis
for useful discussions and comments. This work was partially supported by
grants of the Russian Foundation for Basic Research (RFBR projects Nos
99-01-00091 and 00-15-96691), by INTAS grant No 96-0842 and by CERN--INTAS
grant No 99-0377.

\end{document}